\documentstyle[multicol,aps,epsf]{revtex}

\begin{document}
\input psfig.tex

\title{Winding Angle Distributions for Random Walks and Flux Lines}
\author{Barbara Drossel and Mehran Kardar}
\address{Department of Physics, Massachusetts Institute of  
Technology, Cambridge, Massachusetts 02139}
\date{\today}
\maketitle
\begin{abstract}
We study analytically and numerically the winding of a flux line around 
a columnar defect. Reflecting and absorbing boundary conditions apply to 
marginal or repulsive defects, respectively. In both cases, the winding angle 
distribution decays exponentially for large angles, with a decay constant 
depending only on the boundary condition, but not on microscopic features.
New {\it non-universal} distributions are encountered for {\it chiral} defects
which preferentially twist the flux line in one direction. The resulting 
asymmetric distributions have decay constants that depend on the degree
of chirality. In particular, strong chirality encourages entanglements and
leads to broad distributions. We also examine the windings of flux lines
in the presence of point impurities (random bonds). Our results suggest that
pinning to impurities reduces entanglements, leading to a narrow (Gaussian)
distribution. 

\end{abstract}

\begin{multicols}{2}
\section{Introduction and Summary}
\label{intro}
Winding angles of paths are of great interest not only in mathematics, but 
also in the physics of polymers, and flux lines in high-$T_c$ superconductors.  
The topological constraints produced by the windings of polymers\cite{Gen71} 
or magnetic flux lines\cite{nel88,obu90} around each other, result in 
entangled phases with slow dynamics. The simplest case that can be studied is 
the winding of a 2-dimensional random walk around a point, or equivalently, 
a flux line in 3 dimensions around a columnar pin\cite{nel93}. 
In 1958, Spitzer\cite{spi58} showed that the probability distribution for the 
winding angle $\theta$ of a Brownian path around a point is a Cauchy
law for large time $t$, i.e.
\begin{equation}
\lim_{t\to\infty}p\left(x={2\theta \over\ln t}\right) = {1\over \pi}\, 
{1 \over 1 + x^2}. \label{Cauchy}
\end{equation}
Similar Cauchy laws are obtained for winding around several points in 2 
dimensions\cite{pit86},  and around several straight lines in 3 
dimensions\cite{gal87}. These results are obtained by employing a variety 
of techniques such as standard diffusion equation \cite{spi58,ito65,edw67}, 
path integrals\cite{edw67,wie83}, or probability theory\cite{dur84,pit86,gal87}. 
By contrast, the winding angle of a self-avoiding walk in 2 dimensions obeys 
a Gaussian distribution, the scaling variable being $x=\theta/\sqrt{4\ln t}$\cite{dup88}.
(See also Ref.~\cite{rud88} for an expansion around four dimensions.)

As pointed out in Refs.~\cite{rud87,bel89}, the above Cauchy law has pathological 
properties which make its relevance to any physical situation questionable. 
In particular, because of the slowly decaying tails at large $x$, the 
averages of both $\theta^2$ and $|\theta|$ are infinite. The origin of 
this divergence is that a finite segment of the Brownian walk can wind 
infinitely often around a point center. While this is correct for an 
idealized random walk, in any physical system one expects a cutoff due 
to either finite diameters or stiffness. The case of a Brownian walk in 
2 dimensions around a disc of finite diameter was studied in Ref.~\cite{rud87}. 
The resulting winding angle distribution is \cite{rud87,sal94}
\begin{equation}
\lim_{t\to\infty}p_A\left(x={2\theta \over\ln t}\right) = 
{\pi \over 4 \cosh^2(\pi x/2)}. \label{abs}
\end{equation}
In Ref.~\cite{bel89}, the following result for the winding angle 
distribution for a random walk with steps of finite size is derived 
\begin{equation}
\lim_{t\to\infty}p_R\left(x={2\theta \over\ln t}\right) = {1\over 2}\, 
{1 \over \cosh(\pi x/2)}. \label{refl}
\end{equation}
The same result is obtained in Ref.~\cite{pit86} for the distribution of 
``big windings'' of Brownian motion around two point-like winding centers. 
Saleur\cite{sal94} suggests that the difference between  Eqs.~(\ref{abs}) 
and (\ref{refl}) is due to different boundary conditions for the walkers
at the winding center. A review of many topological and entanglement
properties of polymers can be found in Ref.~\cite{gro93}.

In this paper, we further study the issue of winding angle distributions
and their universality, with particular emphasis on their applicability to
magnetic flux lines (FLs) in high-$T_c$ superconductors. 
We start in Sec.~\ref{theor} by providing a derivation of 
Eqs.~(\ref{Cauchy})--(\ref{refl}) based on conformal properties of random 
walks that does not require any advanced mathematical techniques.
It illustrates well the origin and universality of the exponential tail in
Eqs.~(\ref{abs}) and (\ref{refl}), and explains the factor 2 between their 
two decay constants. We argue that these two cases are applicable 
respectively to the windings of a flux line around a repulsive or
marginal columnar defect.  Actually, most columnar defects are attractive
and localize the flux line to their vicinity. The corresponding probability
distributions (for walkers with initial and final points close to the winding
center) are also calculated in Sec.~\ref{theorD}. 

To test the generality of the analytical results, we performed a number of
numerical tests that are described in Sec.~\ref{numer}. Simulations of
random walks were performed on both square and cubic lattices. 
For reflecting boundary conditions we chose a winding center that
was shifted from the lattice sites crossed by the walker. For absorbing
boundary conditions, the center was one of the lattice sites which the
walkers were not allowed to cross.  A transfer matrix method can be
used to evolve the winding angle distribution with the number of steps $t$.
Despite a rather slow convergence to the asymptotic limit, the numerical
results do indeed support the universality of distributions given in
Eqs.~(\ref{abs}) and (\ref{refl}).

In the course of numerical tests, we encountered one case (reflecting boundary
conditions for a directed path along the diagonal of a cubic lattice) which did 
not conform to any of the above expected universality classes. Further 
examination revealed that we had inadvertently constructed a {\it chiral defect}:
upon encountering the defect, the path had a statistical advantage to wind
in one direction as opposed to the other. In the language of random walks, 
this corresponds to a rotating winding center. The breaking of the symmetry
at the center is in fact a relevant perturbation, and leads to a new type of
probability distribution as discussed in Sec.~\ref{chiral}. Although we do not 
yet have a complete analytic understanding of this novel universality class,
we can account for certain features of this distribution. In particular, 
weak chirality leads to narrow asymmetric distributions, while for strong
chirality both wings of the distribution are widened, i.e. strong chirality of the
defect enhances entanglements.

All the distributions examined in Secs.~\ref{theor}--\ref{chiral} describe the 
windings of {\it ideal} random walks, and in all cases the appropriate scaling 
variable is the combination $x = 2\theta / \ln t$. This universal feature is due 
to the {\it Markovian} nature of these walks, and can be explained as follows: 
After time $t$, the walker has a typical distance $r(t) \propto \sqrt{t}$ from the 
starting point, which is chosen to be close to the winding center. Assuming 
that $r(t)$ is the only relevant length scale, dimensional arguments, combined
with the Markovian property, suggest that $dr/d\theta=r f(\theta)$. The 
rotational invariance of the system implies that $f(\theta)$ must be a constant, 
i.e.  the increase in winding angle cannot depend on the number of windings or
angular  position. Hence,
\begin{equation}\label{idealRW}
d\theta \propto {dr \over r}\propto {dt \over t} = d(\ln t)\, ,
\end{equation}
leading to a scaling variable proportional to $\theta / \ln t$. 

Because of their non-Markovian nature, we cannot apply the arguments
of the preceding paragraph to {\it self-avoiding walks}. Indeed, such walks 
have a Gaussian winding angle distribution in the scaling variable 
$x=\theta/\sqrt{\ln t}$\cite{dup88}. The self-similarity of the walk suggests 
the following scaling argument: Starting from the origin divide the walk into 
segments of 1, 2, $\cdots,\, 2^n\approx t/2$ steps. Since the $\alpha^{\rm th}$ 
segment is at a distance of roughly $2^{\alpha\nu}$ from the center (with 
$\nu = 3/4$) and has a characteristic size of the same order, it is reasonable to 
assume that each segment spans a random angle $\theta_\alpha$ of order one. 
Under the mild assumption that the sum $\theta=\sum_{\alpha=1}^n \theta_\alpha$
satisfies the central limit theorem, we then conclude that $\theta$ is Gaussian
distributed with a variance proportional to $n\propto\ln t$.

We can ask why the above argument of self-similarity does not apply to ideal
walks. The reason is the relevance of the finite winding center. An important 
difference between ideal and self-avoiding walks is that the probability of 
returning to the winding center at the origin asymptotically vanishes for the 
latter because of the exclusion zone set up by self-avoidance. It is thus 
expected that properties of the winding center (size, absorbing versus 
reflecting nature, chirality) are irrelevant for self-avoiding walks. The
ideal walks, on the other hand return to the origin quite often, and upon
rescaling see winding centers of different size. The assumption that
different scaled portions of the ideal walk are self-similar is not correct. 

There is one case where both arguments should hold: An ideal walk 
around a point winding center. The argument based on self-similarity
actually states that the final distribution is obtained from the composition
of $\ln t$ independent random variables. If each variable has a finite
variance, the overall distribution will be Gaussian. If not, other (Levy)
distributions are possible. The Cauchy distribution is in fact a limiting
case for widely distributed variables. The requirement that both 
arguments should hold, immediately selects a Cauchy distribution!

The scaling argument, which by no means is claimed to be exact, should
apply to other self-similar walks where the probability of return to the origin
is small. An interesting example is provided by directed paths in random
media \cite{kardarrev}, which for example describe the behavior of a 
flux line in the presence of (quenched) point impurities. Typical wanderings
of such paths scale as $t^\nu$, with $\nu\approx0.59>1/2$. The pinning
by impurities greatly reduces the probability of the walker returning to the
origin, and the above arguments again suggests that the winding
angle distribution is Gaussian. Although certainly not conclusive,
the numerical results reported in Sec.~\ref{random} appear to support
this prediction. Typical winding angles thus scale as $\sqrt{\ln t}$ as
opposed to $\ln t$ in the pure case.  This simple example thus suggests
that topological entanglements become relatively less important in the
presence of pinning to impurities.

The results of this paper point out the rich behavior already present in
the simplest of problems involving topological defects. Properties of the
winding center (finite size, chirality, $\cdots$), interactions, various
types of randomness are all potentially relevant, leading to different
universal distribution functions. These results could also produce some 
interesting physical manifestations. For example, we demonstrate in 
Sec.~\ref{concl} that there is a sharp crossover between free and coiled  
configurations, if there is an energy proportional to the winding number. 

\section{Derivation of winding angle distributions}
\label{theor}
\subsection{Spitzer's law}
\label{theorA}

Before studying the winding of a Brownian path around a center of finite 
diameter, we first sketch a derivation of Spitzer's law in Eq.~(\ref{Cauchy}). 
We follow the approach in Ref.\cite{dur84}, translating it into a more 
physical  language that does not rely on a familiarity with martingales. 
A basic  ingredient is the invariance of Brownian motion under conformal 
mappings. Let 
\begin{equation}
z(t)=x_1(t)+ix_2(t)
\end{equation}
represent the original walk in the complexified two-dimensional plane. 
The time evolution of each random walker satisfies
\begin{equation}
dz=\eta(t) dt,
\end{equation}
where the complex random velocity has zero mean and is uncorrelated 
at  different times, with 
\begin{equation}
\left\langle \eta(t)\eta^*(t') \right\rangle=2D\delta(t-t').
\end{equation}
The radius of the walker, and its winding angle can be extracted from 
\begin{equation}
\zeta(t)=\ln z(t)=\rho(t)+i\theta(t), 
\end{equation}
where $\rho=\ln r =\ln\sqrt{x_1^2+x_2^2}$.
Since $d\zeta=\eta(t)dt/z(t)$, the stochastic motion of the walker in the 
new complex plane is highly correlated to its location. This feature can 
be removed by defining a new time variable 
\begin{equation}
d\tau={dt\over |z(t)|^2}
\end{equation}
{\it for each walker}, which leads to 
\begin{equation}
d\zeta=\mu(\tau)d\tau, \quad{\rm with}\quad\mu(\tau)=z^*(t)\eta(t). 
\end{equation}
Since 
\begin{equation}
\left\langle \mu(\tau)\mu^*(\tau') \right\rangle=2D|z(t)|^2\delta(t-t')
=2D\delta(\tau-\tau'),
\end{equation}
the evolution
of $\zeta(\tau)$ is that of a Brownian walk. 

For simplicity we choose the initial condition $\zeta(t=\tau=0)=0$, i.e. 
the original walker starts out at $z=1$. We also set the diffusion constant 
to $D = 1/2$, so that the mean square distance over which the walk moves 
during a time $t$ is $\langle r^2(t)\rangle = t$. Since there is a separate 
transformation $\tau(t)$ for each walker, it is not useful to subdivide the 
walkers $\zeta$ according to $\tau$. Instead the walkers of interest are the 
ones that reach lines of fixed $\rho$ {\it for the first time}, independent of the 
value of $\tau$. This is because after a time $t$, the value of $\rho$ is almost 
certainly equal to $\ln t$. In fact, the probability that $r(t)$ is within an 
interval $[\sqrt{\pi} t^{(1 - \epsilon)/2}, \sqrt{\pi} t^{(1 + \epsilon)/2}]$ 
around its mean value of $\sqrt{\pi} t$,
\begin{eqnarray}
p(t,\epsilon) &=& \int_{\sqrt{\pi} t^{(1 - \epsilon)/2}}^{\sqrt{\pi} 
t^{(1 + \epsilon)/2}} {\exp\left(-r^2/2t\right)\over 2\pi t} 2\pi r\, dr 
\nonumber \\
&=& \int_{\sqrt{\pi} t^{-\epsilon}}^{\sqrt{\pi} t^\epsilon} \exp(-s) ds, 
\label{eqepsilon}
\end{eqnarray}
approaches unity in the limit $t \to \infty$. In this limit, the distance 
$r$ from the starting point $z=1$ is identical to the distance from the origin, 
and  $p(t,\epsilon)$ is identical to the probability that $\zeta(\tau)$ is in 
the interval $[0.5(1 - \epsilon)\ln t,0.5(1 +\epsilon)\ln t]$. 

We next shrink the complex plane $\zeta$ by a factor of $(\ln t)/2$. 
The walkers of interest are now those that hit the line with real value of unity 
for the first time. The imaginary coordinate $x$, of the hit is now related to the 
winding angle by $x=2\theta(t)/\ln t$. Thus, in the limit $t\to\infty$, the 
probability $p(x)dx$ is precisely the probability that a Brownian walker 
($2\zeta/\ln t$) starting from the origin, hits a vertical line at a distance one 
from the origin, for the first time, at a height between $x$ and $x + dx$. 

Since the walkers hit the line for the first time at different times $\tau$, 
we need
the probability density $p_1(\tau)$ that the first hit is at $\tau$. 
Its integral
$P_1(\tau)=\int_0^\tau d\tau'p_1(\tau')$, is the probability for the walker
to hit the line at least once before time $\tau$. Since after hitting the wall
the walker is equally likely to proceed in either direction, the latter is 
also twice the probability that the endpoint of the walk is beyond the line 
at $\tau$, and hence given by 
\begin{displaymath}
P_1(\tau)=2\int_{1}^\infty {\exp\left(-r^2/2\tau\right) 
\over \sqrt{2\pi \tau}} dr
= \int_0^\tau  {\exp\left(-1/2s\right)\over\sqrt{2\pi s^3}}  ds \, .
\end{displaymath}
The derivative gives
\begin{displaymath}
p_1(\tau)={d{P_1} \over d\tau}= {1\over\sqrt{2\pi \tau^3}} 
\exp\left(-{1\over 2\tau}\right) \, .
\end{displaymath}
Since the vertical and horizontal components of the walk are independent, 
we finally obtain 
\begin{displaymath} 
p(x) = \int_0^\infty d\tau {\exp\left(-1/2\tau\right) \over 
\sqrt{2\pi\tau^3}} \cdot
{\exp\left(-x^2/2\tau\right) \over \sqrt{2\pi \tau}} = {1\over \pi} 
{1 \over 1 + x^2} \, , 
\end{displaymath}
which is Spitzer's law; exact in the limit $t \to \infty$. 

\subsection{Repulsive defects and finite absorbing centers}
\label{theorB}

The unphysical situation that a path element of finite length can wind 
infinitely often around the origin can be removed by introducing a winding 
center of finite size. A FL  winding around a repulsive columnar defect cannot 
penetrate the defect, and all configurations where the FL and the defect 
intersect have to be removed\cite{rud87}. This situation corresponds to a 
random walk around an absorbing winding center.
Two directed polymers with hard-core interaction winding around each other 
are described by the same situation, the radius $r$ now being the relative 
distance between the two polymers. 

We choose a disc of radius $R<1$ around the origin which cannot be entered 
by the Brownian walk $z(t)$.  A path of length $t$ now has a maximum winding 
angle of $t / R$. The walk is again started from $z(0)=1$, and after a 
long time $t$, its end point is almost certainly at a distance $r$ between 
$\sqrt{\pi} t^{(1-\epsilon)/2}$ and $\sqrt{\pi} t^{(1+\epsilon)/2}$ from the origin, 
as given by Eq.~(\ref{eqepsilon}). (The excluded disc of radius $R\ll\sqrt{t}$ 
has negligble influence in this limit.) The random walks in the complex plane 
$\zeta=\ln z$ consequently have real values in the interval 
$[0.5(1 - \epsilon)\ln t,0.5(1 +\epsilon)\ln t]$ as before. But now the walks  
$\zeta(\tau)$ cannot go  to the left of a line at $a = \ln R$. 
The probability density $p_{A}(x)$ for having a scaled winding angle 
$x = 2\theta/\ln t$ is therefore identical to the probability density that a random 
walk starting at the origin hits a line at distance one, for the first time, at height 
$x$ without going beyond a line at distance $2|\ln R|/\ln t$ on the opposite side. 

Let $P_{ a, b}(y,\tau)$ denote the probability  that a Brownian walk
starting at $y \in [ a, b]$ hits the point $ b$ for the first time during a time 
interval $\tau$, without hitting point $a$ previously. If we consider the points 
$a$ and $b$ as absorbing boundaries, this is identical to the probability 
that the walk is absorbed at boundary $b$ before time $\tau$. Since for 
sufficiently small $\Delta \tau$, the walker is only a short distance
$\Delta y$ from its starting point, we have
\begin{eqnarray*}
 && P_{ a, b}(y,\tau) = \int_0^\infty d(\Delta y) {1\over \sqrt{2\pi\Delta 
\tau}}
\exp\left[-{(\Delta y)^2\over 2\Delta \tau} \right]  \\
& &\times 
\left[P_{ a, b}(y + \Delta y,\tau - \Delta \tau) + P_{ a, b}(y - \Delta y,
\tau - \Delta \tau) \right] \, .
\end{eqnarray*}
Expanding the above equation to the order of $\Delta t$ indicates that  
$P_{ a, b}(y,\tau)$
satisfies a diffusion equation. The appropriate boundary conditions are
$P_{ a, b}( a,\tau) = 0$ and $P_{ a, b}( b,\tau) = 1$ with the initial value 
$P_{ a, b}(y,0) = 0$, resulting in\cite{kni81}
\begin{eqnarray*}
P_{ a, b}(y,\tau) &=& {y -  a \over  b -  a} + {2\over \pi} 
\sum_{\nu = 1}^\infty {(-1)^{\nu + 1} \over \nu} \sin\left({\pi 
\nu (y -  a) \over  b -  a}\right) \\
& & \quad \times
\exp\left[-{1\over 2} \left({\pi \nu \over  b -  a}\right)^2 \tau\right]\, . 
\end{eqnarray*}
The probability  that the walk is absorbed at the right-hand boundary
during the time interval $[\tau,\tau + d\tau]$, is 
$ d\tau \partial_\tau P_{ a, b}(y,\tau) $. 
Note, however, that
$\int_0^\infty d\tau \partial_\tau P_{ a, b}(y,\tau) =(y-a)/(b-a)$, i.e. equal 
to the total fraction of particles absorbed at the right-hand boundary 
(inversely proportional to the separations from the boundaries). 
To calculate $p_A(x)$, we need the fraction of these walks absorbed 
between $\tau$ and $\tau+d\tau$, equal to $((b-a)/(y-a)) \partial_\tau P$. Hence
(with $ a = 2\ln R/ \ln t$, $ b = 1$ and $y = 0$)
\begin{eqnarray}
p_{A}(x) & = & \int_0^\infty d\tau {1 -  a\over - a}{\partial 
P_{ a,1}(0,\tau) \over \partial \tau}{\exp\left(-x^2/2\tau\right) \over 
\sqrt{2\pi\tau}}   \nonumber \\
&=& \int_0^\infty d\tau \sum_{\nu = 1}^\infty {(-1)^{\nu + 1} \over 
\sqrt{2\pi\tau}}  {\pi \nu \over  a(1 -  a)} \sin\left( {\pi \nu  
a\over 1 -  a}\right)\nonumber \\
& & \quad \times \exp\left[-{1 \over 2} \left( {\pi \nu \over 1 -  a}\right)^2
 \tau - {x^2 \over 2\tau}\right] \nonumber \\
&=& \sum_{\nu = 1}^\infty {(-1)^{\nu + 1} \over  a} \sin\left( 
{\pi \nu  a\over 1 -  a}\right) \exp\left[-{\pi\nu |x| \over (1 -  a)}
\right] \, .\nonumber 
\end{eqnarray}
The last step is achieved by first performing a Fourier transform with 
respect to $x$, followed by integrating over $\tau$, and finally inverting 
the Fourier transform. (Alternatively, the $\tau$ integration can be performed 
by the saddle point method.) In the limit of large $t$, the variable $ a$ is very 
small, and we can replace the sine--function by its argument. Taking the sum 
over $\nu$, we find
\begin{equation}
p_{A}(x) = {\pi \over  1 -  a} {\exp\left[\pi x/(1 -  a)\right] 
\over \left\{\exp\left[\pi x/(1 -  a)\right] + 1\right\}^2}\, . \label{eqR}
\end{equation}
Changing the variable from $x$ to 
\begin{displaymath}
\tilde x = {x\over (1 -  a)} = {2\theta\over \ln\left(t/R^2\right)}\, ,
\end{displaymath}
and noting that $p_{A}(x) dx = p_A(\tilde x) d\tilde x$, leads from 
Eq.~(\ref{eqR})  to
\begin{equation}
p_A\left(\tilde x = {2\theta\over \ln\left(t/R^2\right)}\right) 
= {\pi \over 4 \cosh^2(\pi\tilde x/2)}\, .
\label{abs2}
\end{equation}

This is identical to Eq.~(\ref{abs}), since $\tilde x = x$ for large $t$. 
The use of the variable $\tilde x $ is, however, more appropriate, 
since it makes the argument of the logarithm dimensionless, and 
explicitly indicates the basic time unit. The above distribution, 
which is exact in the limit $t \to \infty$, has an exponential decay 
for large $\tilde x$, in contrast to Spitzer's law in Eq.~(\ref{Cauchy}). 
In particular, the mean winding angle is now finite. The analogy to 
random walkers in the plane $\zeta$, confined by the two walls, 
provides simple physical justifications for the behavior of the 
winding angle. In the presence of both walls, the diffusing
particle is confined to a strip, and loses any memory of  its starting 
position at long times. The probability that a particle that has already 
traveled a distance $\theta$ in the vertical direction proceeds a 
further distance $d\theta$ without hitting either wall is thus 
independent of $\theta$, leading to the exponential decay. 
On the other hand, if there is no confining wall on the left-hand side, 
the particles may diffuse arbitrarily far in that direction, making it less 
probable to hit the wall on the right hand side. The absence of a 
characteristic confining length thus leads to the unphysical 
divergence in Spitzer's law. 

We also expect that different shapes of the winding center  do not modify 
Eq.~(\ref{abs2}). This is because of the division by $\ln t/2$, which maps 
any excluded shape to the vertical line at $a=0$ for long enough times.
However, different shapes should result  in different effective values for 
$R$, and the time constant in Eq.~(\ref{abs2}). For long times, the polar 
angle of the particle may take any value between $0$ and $2\pi$ with equal 
probability, and the particle sees an averaged radius of the winding center. 

The crossover to Spitzer's law in the limit $R\to0$ is not apparent from the
above expression for $p_A(x)$. The transition can be made in the initial 
equation for $p_{A}(x)$, but not in the final result of Eq.~(\ref{abs2}), where 
time is measured in units of $R^2$. The time scale in Spitzer's law is set by 
the distance of the initial point from the winding center, which is infinite in 
units of the radius of the winding center. Consequently one time unit in 
Eq.~(\ref{Cauchy}) corresponds to infinitely many time units in 
Eq.~(\ref{abs2}), and the limits $t \to \infty$ or $R \to 0$ or $\tilde x \to 0$ 
in  Eq.~(\ref{abs2}) correspond to the limit $t \to 0$ or $x \to \infty$ 
in Eq.~(\ref{Cauchy}).

\subsection{Neutral defects and reflecting winding centers}
\label{theorC} 

For windings around several point centers, or the winding of a random 
walk on a lattice around a point different from the vertices of the 
lattice\cite{pit86,bel89}, no configurations are forbidden, and no walks are 
removed. The resulting winding angle distribution in both cases is given 
by Eq.~(\ref{refl}). The random walk on a lattice can be regarded as a
model for a directed polymer of finite stiffness, the step size being of the 
order of the persistence length. This situation applies to magnetic FLs 
winding around a columnar defect that is marginal, i.e. neither attractive 
nor repulsive. We can obtain  Eq.~(\ref{refl}) by repeating the calculations 
of the previous subsection, but replacing the absorbing boundary 
condition $P_{ a, b}( a,\tau) = 0$ with the reflecting condition 
$\partial P_{ a, b}( y,\tau)/\partial y\mid_{y=a} = 0$. There is thus no 
current leaving the system at point $a$, and walkers which hit the 
winding center are reflected. We then find
\begin{eqnarray}
P_{a,b}(y,\tau) &=& 1 - {2\over \pi} \sum_{\nu = 0}^\infty {1\over 
\nu + 1/2} \sin\left(\pi ( \nu + 1/2) (b-y)\over b-a \right) \nonumber\\
&&\qquad \times \exp\left[-{1\over 2} \left({\pi (\nu + 1/2) \over  b -  a}
\right)^2 \tau\right] \label{reflP}
\end{eqnarray}
and
\begin{eqnarray}
p_{R}(x) & = & \int_0^\infty d\tau {\partial P_{ a,1}(0,\tau) \over \partial 
\tau}{\exp\left(-x^2/2\tau\right) \over \sqrt{2\pi\tau}}   \nonumber \\
&=& \sum_{\nu = 0}^\infty {1 \over  1 - a} 
\sin\left( {\pi (\nu + 1/2) \over 1 -  a}\right)\nonumber\\
&&\qquad \times  \exp\left[-{\pi(\nu+1/2) |x| \over (1 -  a)}\right] 
\nonumber \\
&=& {1\over 2 \cosh(\pi\tilde x /2)}\,,  \label{refl2}
\end{eqnarray}
where again time is measured in units of $R^2$, and the limit 
$t \to \infty$ has been taken. 
For large $\tilde x$, where the walk has lost the memory of its initial 
distance from both walls, this probability decays exponentially as
$\exp\left[-\pi\tilde{x}/2\right]$. A random walk confined between an 
absorbing and a reflecting wall that have a distance 1 can be mapped 
to a random walk confined between two absorbing walls at distance 2. 
After rescaling the wall distance and the $\tilde x$-coordinate by two, 
this explains the factor 1/2 between the decay constants in the tails 
of the distributions in Eqs.~(\ref{abs}) and (\ref{refl}). 

\subsection{Attractive defects and returning walks}
\label{theorD}

A magnetic FL winding around an attractive columnar defect of radius $b_0$ 
and binding energy $U_0$ per unit length, is bound to that defect. If the 
temperature is above a crossover temperature $T^* \propto b_0 \sqrt{U_0}$, 
the line is only weakly bound and wanders horizontally over the distance of 
the localization length $l_{\perp}(t) \simeq b_0 \exp[(T/T^*)^2]$\cite{nel93}. 
The mean vertical distance $l_z$ between consecutive intersections of the 
FL with the defect is consequently proportional to $l_{\perp}^2$. Over this 
distance the FL can be approximated by a directed walk which returns to its 
starting point (the winding center) after a time $l_z$. 

Using the result in Eq.~(\ref{abs2}), we can derive the winding angle 
distribution $p_A^o(\tilde x)$ for such confined random walks. Each walk that 
returns to its starting point after time $t$ (in the $z$-plane) is composed of two 
walks of length $t/2$ going from the starting point to $z(t/2)$. As we have seen 
in subsection \ref{theorA}, almost all walks of length $t/2$, when mapped on 
the plane $2\zeta/\ln(t/2)$, have their endpoint on a vertical wall at distance 
1 from the origin. The winding angle distribution for these walks is given 
by Eq.~(\ref{abs2}), with $t$ replaced by $t/2$. The probability that a walk 
that returns to its starting point has a winding angle $\theta$ is therefore 
obtained by adding the probabilities of all combinations of two walks of length 
$t/2$ whose winding angles add up to $\theta$, i.e.
\begin{eqnarray}
p_A^o(\tilde x) &=& \int_{-\infty}^\infty dy {\pi\over 4 \cosh^2(\pi y)}
{\pi\over 4 \cosh^2\left(\pi(\tilde x -  y )\right)}\nonumber\\
&=& {\pi\over2 \sinh^2(\pi\tilde x/2)}\left({\pi \tilde{x} \over 2}
\coth({\pi \tilde{x} \over 2}) - 1\right) \,, \label{abso}
\end{eqnarray}
where $\tilde x = 2\theta / \ln(t/2R^2)$.  Surprisingly, this result is different
from the one given in Eq.~(4.1) of Ref.~\cite{sal94}. We are not sure of the 
origin of this discrepancy,  but note that the result in  
Ref.~\cite{sal94}  has a divergence at  $\tilde x = 0$ which is unphysical. 

If in place of convoluting two absorbing probability distributions, as in
Eq.~(\ref{abso}), we use the probability distributions for reflecting boundary 
conditions, the final result is modified to
\begin{equation}
p_R^o(\tilde x) = {\tilde{x} \over 2}\,{1\over \sinh(\pi\tilde x/2)}\,. \label{reflo}
\end{equation}
For large $\tilde{x}\approx x$, the above two expressions decay as
$x\exp\left[-\pi x\right]$ and $x\exp\left[-\pi x/2\right]$ respectively.
A FL of length $T$ is roughly broken up into $T/l_z$ segments between
contacts with the attractive columnar defect. We can assume that the winding 
angle of each segment is independently taken from the probability distribution 
in Eq.~(\ref{abso}) with $t\approx l_z$. Adding the winding angle distributions 
of all segments leads a Gaussian distribution centered around $\tilde x = 0$, 
and with a variance proportional to $T\ln(l_z)/l_z$. 

\section{Universality of the winding angle distribution}
\label{numer}

At first sight it may seem surprising that the winding angle distribution in
Eq.~(\ref{refl2}) for random walks around a reflecting center of 
finite diameter is identical to the distribution of ``large windings'' for 
Brownian motion around two point-like centers\cite{pit86}, or to
distribution for random walks of finite step size\cite{bel89}. 
The increase in winding angle is largest when the walk is close to the 
winding center, and should thus be quite sensitive to the lattice structure, 
or the ratio between step size and winding center size.  However, 
a closer examination of the random walks in the scaled $\zeta$-plane
reveals that, in the limit $t \to \infty$, almost no increase in the variable 
$\tilde x$ occurs when the walker is within a certain finite distance from 
the winding center (corresponding to an infinitely small distance from the 
reflecting wall in the scaled $\zeta$-plane). Hence, microscopic details  
do not enter the winding angle distribution, and Eq.~(\ref{refl2}) 
holds as long as there is a winding center of nonvanishing size (or, 
equivalently, an upper cutoff to the maximum possible winding angle per 
unit time), and as long as no walk coming close to the winding center is 
absorbed.

If a walk hitting the winding center is absorbed with a nonvanishing 
probability, almost all walks are absorbed in the limit $t \to \infty$, 
and the winding angle distribution of the surviving walks is given by 
Eq.~(\ref{abs2}). As for reflecting boundary conditions, microscopic details 
of the system do not influence this distribution. Apart from these two 
universality classes, we shall later encounter a third class of walks 
corresponding to random motion around a rotating winding center; 
the winding angle distribution in this case depends on the rotation speed 
of the center. In this section, we check numerically the former situations, 
where the symmetry between $\pm\theta$ is not broken. 

We performed computer simulations of random walks on a square lattice with 
both reflecting and absorbing boundary conditions. Reflecting boundary 
conditions are simply realized by choosing a winding center different from 
the vertices of the lattice, and thus never crossed by the walker. 
On the other hand, to simulate absorbing boundary conditions, the winding
center  is chosen as one of the lattice sites (say the origin), 
but no walk is allowed to go through this point. 

The winding angle distributions are most readily obtained using a transfer 
matrix method which calculates the number of all walks with given winding 
angle and given endpoint after $t$ steps, from the same information after 
$t-1$ steps. The winding center is at $(0.5,0.5)$ for reflecting boundary 
conditions, and at the origin for absorbing boundary conditions.
The walker starts at (1,0), and the winding angle is increased or 
decreased by $2\pi$ every time it crosses the positive branch
 of the $x_1$--axis. 
Due to limitations in computer memory, 
we applied a cutoff in system size and winding angle for times $t > 120$, 
making sure that the results are not affected by this approximation. The largest 
times used, $t=9728$, required approximately 3 days to run on a Silicon 
Graphics Indy Workstation.

\begin{figure}
\narrowtext
\vskip -1cm
\centerline{{\epsfysize=3in 
\epsffile{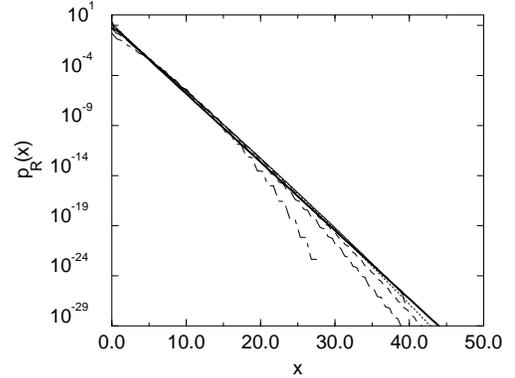}}}
\vskip -1cm
\caption{Winding angle distribution for random walks on a square lattice 
with {\it reflecting} boundary conditions for $t = 38$ (dot-dashed), 
152 (long dashed),  608 (dashed), 2432 (dotted), and 9728 (solid). 
The horizontal axis is $x=2\theta/\ln(2t)$. The thick solid line is the 
analytical  result of Eq.~(\protect\ref{refl}). 
}
\label{squarerefl}
\end{figure}

\begin{figure}
\narrowtext
\vskip -1cm
\centerline{{\epsfysize=3in 
\epsffile{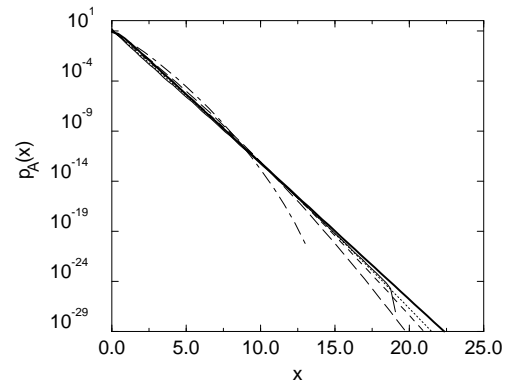}}}
\vskip -1cm
\caption{Winding angle distribution for random walks on a square lattice 
with {\it absorbing} boundary conditions. The symbols and the variable 
$x$ are the same as in the previous figure. The thick solid line is the 
analytical result of Eq.~(\protect\ref{abs}).
}
\label{squareabs}
\end{figure}

Figures \ref{squarerefl} and 
\ref{squareabs} show the simulation results for the two cases. The asymptotic 
exponential tails predicted by theory can clearly be seen; deviations from 
the theoretical curve for smaller values of the scaling variable $x = 2\theta/\ln(2t)$ 
are due to the slow convergence to the asymptotic limit.  Since the scaling 
variable depends logarithmically on time, the asymptotic limit is reached only 
for large $\ln t$. Note that the only free parameter in fitting to the analytical form
is the characteristic time scale appearing inside the logarithm. With $t$ 
measured in units of single steps on the lattice, we found that a factor 2 in the 
scaling variable provides the best fit. In the limit $t \to \infty$, different scales of
$t$ give of course the same asymptotic winding angle distribution. 

\begin{figure}
\narrowtext
\vskip -1cm
\centerline{{\epsfysize=3in 
\epsffile{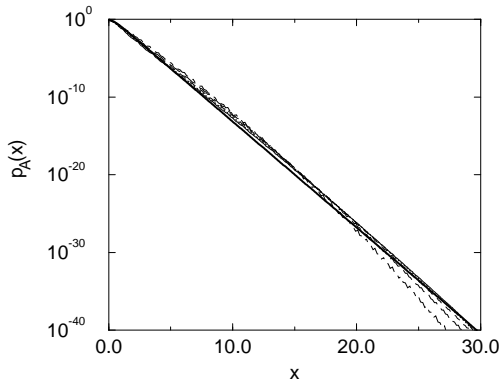}}}
\vskip -1cm
\caption{Winding angle distribution around the preferred direction for a 
flux line (directed path) in 3 dimensions for $t = 243$ (dot-dashed), 729 
(long dashed),  2187 (dashed), 6561 (dotted),  and 19684 (solid). 
The horizontal axis is $x=2\theta/\ln(2t)$. The thick solid line is the 
analytical result of Eq.~(\protect\ref{abs}).
}
\label{fig1}
\end{figure}

To further test the universality of these distributions, we also performed 
simulations of a  flux line (directed path) proceeding along the diagonal 
of a cubic lattice in three dimensions. The FL starts at $(1,0,0)$, and at each
step increases one of its three coordinates by 1. We determined the 
winding angle distribution around the diagonal $(1,1,1)$-- direction, 
excluding from the walk all points which are on this diagonal (a repulsive
columnar defect, corresponding to the case of absorbing winding center).
The excluded points lie on the origin when the FL is projected in a plane 
perpendicular the diagonal. A cutoff of $243$ in system size was imposed for 
the transfer matrix calculations. The winding angle distribution 
$p(x = 2\theta/\ln t)$ is shown in Fig.~\ref{fig1} for different times. As for 
the square lattice, an exponential tail with decay constant of $\pi$ can 
be clearly seen. Our numerical results, as well as the analytical considerations, 
thus indicate clearly that the winding angle distributions for reflecting and 
absorbing boundary conditions are universal and do not depend on 
microscopic details.

Due to the special properties of directed paths along the diagonal of the cube, 
the case of reflecting boundary conditions leads to an asymmetry between 
windings in positive and negative directions. This is because it takes only 
three steps to make the smallest possible winding in one direction, but six 
steps in the opposite direction. This situation is discussed in detail in 
Sec.~\ref{chiral}. 

\section{Chiral defects and rotating winding centers}
\label{chiral}

\begin{figure}
\narrowtext
\vskip -0.5cm
\centerline{{\epsfysize=3in 
\epsffile{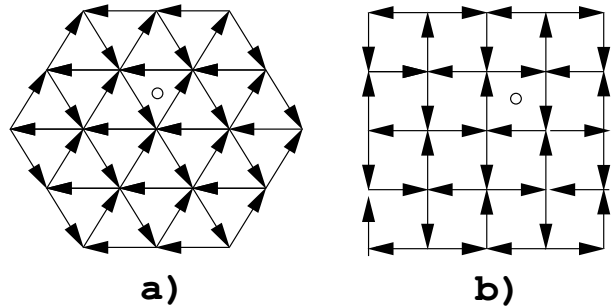}}}
\vskip -2cm
\caption{Triangular and square lattices with directed bonds.
The winding centers are indicated by a circle o.
}
\label{gitter}
\end{figure}

So far, we only considered situations that are symmetric with respect to the angles 
$\pm\theta$. For directed paths on certain lattices, however, this symmetry is 
broken, as mentioned in Sec.~\ref{numer}. A directed walk that proceeds at each 
step along $+x_1$, $+x_2$, or $+x_3$ direction on a cubic lattice can be mapped on a 
random walker on a two-dimensional triangular lattice as indicated in figure 
\ref{gitter}(a). 
Each bond can be crossed only in one direction,  and the winding center for 
reflecting boundary conditions must be different from the vertices of the 
lattice. It is apparent from this figure that the random walker can go around 
the center in 3 steps in one angular direction, but in no less than 6 steps 
in the other direction.

An alternative description is obtained by examining the position of the walker
after every three time steps. The resulting coarse-grained random walk takes 
place on a regular triangular lattice, but now the walker has a finite 
probability of
$3\times 2/3^3=2/9$ of staying at the same site. If this site is one of the three 
points next to the winding center, the winding angle is increased by $2\pi$
in one of the six possible configurations that return to the site after three steps. 
In other words, the walker has a finite probability of having its winding angle 
increased in the proximity of the center. The amount of this biased increase in 
angle depends on the structure of the lattice and will be different 
for other directed lattices. An equivalent physical situation occurs 
for Brownian motion around a rotating winding center, e.g. a rotating reflecting 
disc which does not set the surrounding gas or liquid into motion. 

Since this angular symmetry breaking is already present in the above simple
example of a directed walk, it is also likely to occur in more realistic physical 
systems, such as with screw dislocations. 
We thus use the term {\it chiral} columnar defect, to indicate that
each time the FL comes close to the defect, it finds it easier to wind around
in one direction as opposed to the other. (Of course, to respect the reflecting
boundary conditions, there must be no additional interaction with the defect.)

After mapping to the rescaled $\zeta$-plane introduced in Sec.~\ref{theor},
the above situations can be modeled by an upward moving reflecting wall 
on the vertical axis. Each time a Brownian walker hits this wall, its vertical 
position $x = 2\theta/\ln t$ is increased by a small amount $2\Delta\theta/\ln t$.
Let us now determine the net shift in $x$ due to the motion of the wall for a
walker that survives for a time $\tau$ in the rescaled $\zeta$-plane, before it is 
absorbed at the right-hand wall (recall that $t$ is the time in the original system, 
while $\tau$ refers to the time in the rescaled $\zeta-$plane, after the conformal 
mapping).  

To obtain the full solution, it is necessary to solve the two-dimensional diffusion
equation with moving boundary conditions. Since we are mainly interested in
the exponential tails of the winding angle distribution, we restrict our analysis
to the limit of large times $\tau$, and determine the increase in $x$ due to
encounters with the reflecting moving wall in this limit. A Brownian walker 
that has survived for a sufficiently long time $\tau$ forgets its initial horizontal
position. The mean number of encounters with the reflecting wall, and 
consequently the increase in $x$ due to the motion of the wall, is then 
expected to be simply proportional to the considered time interval. Assuming
the validity of the central limit theorem in the limit $\tau \to \infty$, the 
probability distribution of this increase $\Delta x$ in $x$ is given by
\begin{equation}\label{pDelta}
p_\Delta(\Delta x)={1\over\sqrt{2\pi\beta^2\tau}}
\exp\left[-{(\Delta x -\alpha \tau)^2\over 2\beta^2 \tau }\right].
\end{equation}
The parameters $\alpha$ and $\beta$ are related to the velocity of the wall
(chirality of the defect) by $\alpha\propto\beta\propto v$. Presumably
Eq.~(\ref{pDelta}) can be obtained directly from properties of random walks,
providing the exact coefficients in the above proportionality.

We can now modify Eq.~(\ref{refl2}) to
\begin{eqnarray}
p^c_{R}(x) & = & \int_0^\infty d\tau \int_{-\infty}^{\infty} d(\Delta x)
p_\Delta(\Delta x){\partial P_{ a,1}(0,\tau) \over \partial \tau} \nonumber \\
&&\qquad \times {1 \over \sqrt{2\pi\tau}}
\exp\left[-{(x-\Delta x)^2\over 2 \tau}\right] \nonumber \\
 & = &  \int_0^\infty d\tau 
{\partial P_{ a,1}(0,\tau) \over \partial \tau} \nonumber\\
&&\qquad \times {1\over \sqrt{2\pi\tau(1+\beta^2)}}
\exp\left[-{(x-\alpha \tau)^2\over 2\tau(1+\beta^2)}\right] .\nonumber \\
\nonumber
\end{eqnarray}
In the limit of large $x$ it is sufficient to take the first term in the series 
expansion for $P_{ a,1}(0,\tau)$ given in Eq.~(\ref{reflP}), leading to
\begin{equation}\label{chir}
p^c_{R}(x) = \exp\left[{\alpha x\over 1+\beta^2} 
-{|x|\over 1+\beta^2}\sqrt{\alpha^2+{\pi^2\over 4}
\left(1+\beta^2\right)}\right]. 
\end{equation}
The effect of the moving wall on the winding angle distribution is thus a 
systematic shift in the slopes of the exponential tails. For small values of
chirality the slopes on the two sides are changed to $\pi/2\pm \alpha$. 
Due to the velocity dependence, these asymmetric distributions are 
clearly non-universal. At large chiralities the slopes vanish as 
$\alpha^2/\beta$ resulting in quite wide distributions. Apparently strong
chirality of a defect increases the probability of entanglements.

\begin{figure}
\narrowtext
\vskip -1cm
\centerline{{\epsfysize=3in 
\epsffile{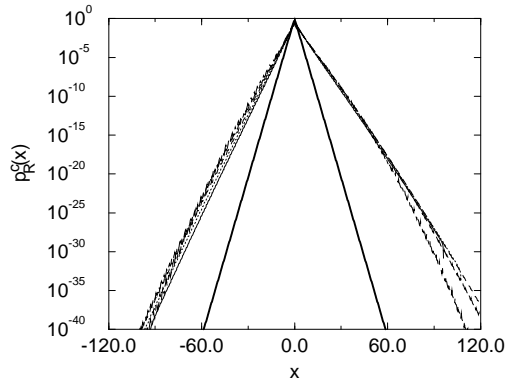}}}
\vskip -1cm
\caption{Winding angle distribution around the preferred direction for a 
random walk on a directed triangular lattice for $t = 243$ (dot-dashed), 729 
(long dashed),  2187 (dashed), 6561 (dotted), and 19684 (solid). The 
scaling variable is $x=2\theta/\ln(2t)$. The thick solid line is the 
distribution  given in Eq.~(\protect\ref{refl}).
}
\label{chiral1}
\end{figure}

\begin{figure}
\narrowtext
\vskip -1cm
\centerline{{\epsfysize=3in 
\epsffile{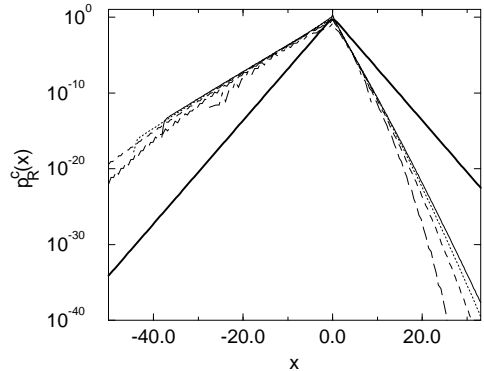}}}
\vskip -1cm
\caption{Winding angle distribution for a random walk on a directed square 
lattice for  $t = 38$ (dot-dashed), 152 (long dashed),  608 (dashed), 2432 
(dotted), and 9728 (solid).
The scaling variable is $x=2\theta/\ln(2t)$. The thick solid line is the 
distribution Eq.~(\protect\ref{refl}).
}
\label{chiral2}
\end{figure}

Fig.~\ref{chiral1} shows our simulation results for the winding angle 
distribution for a walk on the above mentioned directed triangular lattice. 
The asymmetry due to the shift is clearly visible, and the winding angle 
distribution is wider than for a stationary wall. This case thus exemplifies
the strong chirality limit discussed in the previous paragraph.
We also simulated a square lattice with directed bonds as indicated in
Fig.~\ref{gitter}(b). The corresponding winding angle distribution is shown in
Fig.~\ref{chiral2}. The distribution is again asymmetric, but not as wide 
as in the previous one, and more similar to that expected in the weak
chirality regime.

\section{Winding angles in random media}
\label{random}

FLs in high-$T_c$ superconductors are pinned by oxygen impurities. Such
pinning is quite essential to enhancing the current carrying capacity of the
superconductor in its mixed phase\cite{HTSC}. Actually, the pinning to the
(quenched) random impurities fundamentally modifies the properties of FLs,
leading to glassy phases. The simplest example is again a single FL as
discussed in the previous section. The behavior of the FL in the presence
of point impurities can be modeled by a directed path on a lattice with random 
bond energies\cite{kardarrev}. In 3 or less dimensions, the line is always pinned at 
sufficiently long length scales. An important consequence of the pinning is that the 
path wanders away from the origin much more than a random walk, its transverse 
fluctuations scaling as $t^\nu$, where $\nu\approx 0.62$ in three dimensions, 
and $\nu=2/3$ in two dimensions\cite{AmarFamily,KimBrayMoore}.
The probability of such paths returning to the winding center are thus greatly
reduced, and thus the winding probability distribution is expected to change. 

We examined numerically the windings of a directed path along the diagonal
of a cubic lattice (see Fig.~\ref{gitter}(a)). To each bond of this lattice was assigned
an energy randomly chosen between 0 and 1. Since the statistical properties 
of the pinned path are the same at finite and zero temperatures, we determined
the winding angle of the path of minimal energy by a transfer matrix method.
For each realization of randomness this method\cite{kardarrev} finds the
minimum energy of all paths terminating at different points, and with different
winding numbers. This information is then updated from one time step to the
next. From each realization we thus extract an optimal angle as a function of
$t$. The probability distribution is then constructed by examining 2700 different
realizations of randomness. To improve the statistics, we averaged over positive 
and negative winding angles. 

\begin{figure}
\narrowtext
\vskip -1cm
\centerline{\epsfysize=3in
\epsffile{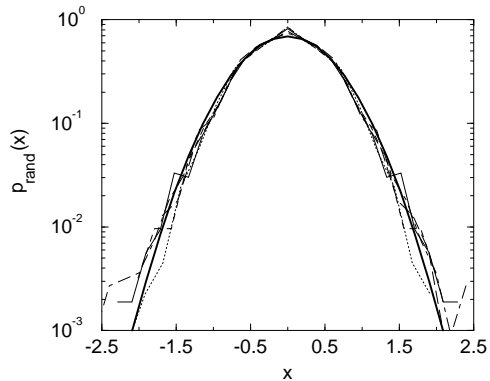}}
\vskip -1cm
\caption{Winding angle distribution for a directed path in a random
3-dimensional system for  $t = 120$ (dotted), 240 (dashed),
480 (long dashed), 960 (dot-dashed),  and 1920 (solid). 
The thick solid line is the  Gaussian distribution in Eq.~(\protect\ref{gauss}).
}
\label{fig3a}
\end{figure}

\begin{figure}
\narrowtext
\vskip -1cm
\centerline{{\epsfysize=3in 
\epsffile{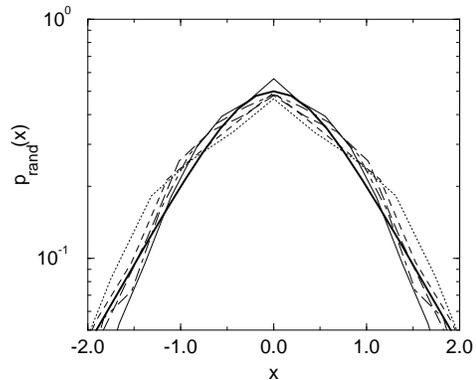}}}
\vskip -1cm
\caption{Same distribution as in previous figure, but with $x=2\theta/\ln t$. 
The thick dotted line is the distribution for the pure case given in Eq.~(\protect\ref{abs}).
}
\label{fig3b}
\end{figure}

The resulting distributions are shown in figures \ref{fig3a} and \ref{fig3b}.  The 
scaling variable in Fig.~\ref{fig3b} is $2\theta/\ln t$ and the results are compared 
to Eq.~(\ref{abs}) that is expected in the pure system. A much better fit is achieved 
with the scaling variable $x = \theta/2\sqrt{\ln t}$ as indicated in Fig.~\ref{fig3a}. 
This scaling form is motivated by that of self-avoiding walks, which in two 
dimensions follow a Gaussian distribution
\begin{equation}\label{pSAW}
p_{SA}\left(x = {\theta\over 2\sqrt{\ln t}}\right)= 
{1\over \sqrt{\pi}}\exp\left(-x^2\right). 
\end{equation}
The result of the data collapse in Fig.~\ref{fig3a} agrees well with 
the Gaussian distribution  
\begin{equation}
p_{\rm rand}\left( x= {\theta\over 2\sqrt{\ln t} }\right)=
\sqrt{1.5\over \pi}\exp\left(-1.5x^2\right) \, . 
\label{gauss}
\end{equation}

Directed paths in random media and self-avoiding walks share a number of
features which make the similarity in their winding angle distribution plausible.
Both walks meander away with an exponent larger than the random walk 
value of 1/2. (The exponent of 3/4 for self-avoiding walks is larger than 
$\nu\approx 0.59$ for FLs in 3 dimensions.) As a result, the probability
of returns to the origin is vanishingly small in the $t\to\infty$ for both
paths, and thus the properties of the winding center are expected to be
irrelevant. (A simple scaling argument suggests that the number of returns
to the origin scale as $N(t)\propto1/t^{1-2\nu}$.)
The conformal mapping of section \ref{theor} cannot be 
applied in either case: The density and size of impurities in a random medium
becomes coordinate dependent under this mapping, as does the excluded 
volume effect. The winding angle distribution for self-avoiding walks in
Eq.~(\ref{pSAW}) has been 
calculated using a more sophisticated mapping\cite{dup88,sal94}. 
As a similar exact solution is not currently available for FLs in random media,
we resort to the scaling argument presented next.

Let us divide the self-avoiding walk, or the directed path, in segments going from 
$t/2$ to $t$, from $t/4$ to $t/2$, etc., down to some cutoff length of the order of the 
lattice spacing, resulting in a total number of segments of the order of $\ln t$. 
The statistical self-similarity of the walks suggests that a segment of length 
$t/2^n$ can be mapped onto a segment of length $t/2^{n+1}$ after rescaling by 
a factor of $1/2^\nu$. Under this rescaling, the winding angle is (statistically 
speaking) conserved, and consequently all segments have the same winding 
angle distribution. Convoluting the winding angle distributions of all segments, 
and assuming that the correlations between segments do not invalidate the 
applicability of the central limit theorem, leads to a Gaussian distribution with 
a width proportional to $\ln t$. This argument does not work for the random walks 
considered in section \ref{theorB}--\ref{theorD}, since the finite radius of the
winding center is a relevant parameter. Different segments of the walk are
therefore not statistically equivalent, as they see a winding center of different 
radius after rescaling. 

Another interesting quantitity to study is the difference in free energy 
between configurations of different winding numbers. This provides an estimate
of the free energy that can be gained (or lost) by a FL upon changing its degree
of entanglement. In the pure system, this quantity can be obtained directly from
the winding angle distribution. For large $t$, the difference in free energy between 
configurations of winding number $n$ and 0 is calculated by expanding
Eq.~(\ref{abs}) for small $x=4\pi n/\ln t$, and is given by
\begin{equation}\label{Df}
F(n) - F(0) = k_BT {2\pi^4 n^2 \over (\ln t)^2}.
\end{equation}
(The corresponding result for absorbing boundary conditions is larger by a factor 2.)

\begin{figure}
\narrowtext
\vskip -1cm
\centerline{{\epsfysize=3in 
\epsffile{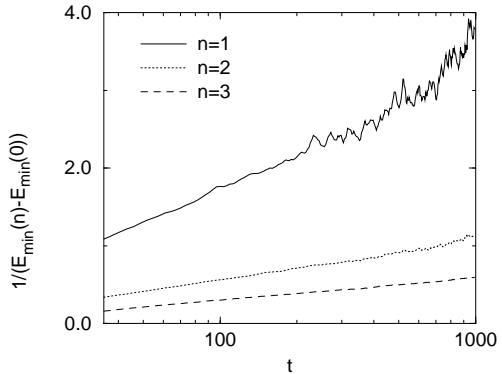}}}
\vskip -1cm
\caption{Difference in minimal energy between directed polymers of different 
winding numbers, averaged over 1019 realizations of randomness. 
}
\label{fig4}
\end{figure}

In the presence of quenched randomness, there exists no obvious relation 
between the free energy and the entropy, and we therefore determined the 
difference between energies of minimal paths of different winding numbers.
The simulation results are depicted in Fig.~\ref{fig4}. Somewhat surprisingly,
the difference in energy vanishes in the large $t$ limit, as in Eq.~(\ref{Df}).
Naive arguments may have suggested that the energy cost associated with 
changing the winding number either saturates at a finite value, or possibly
even grows as $t^{2\nu-1}$ as in typical energy fluctuations \cite{kardarrev}.
Nonetheless, Fig.~\ref{fig4}  suggests a $1/\ln t$ dependence on length, 
although we cannot rule out some larger power of $1/\ln t$. The slopes of
the curves in this figure have ratios of approximately 1/3,  1/6, and 1/2, 
different from 1/4, 1/9, and 4/9 for an  $n^2$ dependence.  We could not
collapse the data using the scaling variable $n/\sqrt{\ln t}$ as in Fig.~\ref{fig3a}.

Similar results are expected to hold for the windings and entanglements
of two FLs around each other.  The interactions between the FLs are small
at large separations (very small density), and the relative distance between
the lines behaves in the same way as the separation of one line from a
columnar defect. In the pure system, we had to distinguish between
repulsive, neutral, and attractive defects. Since in the presence of point
impurities the line does not return to the origin as often, the interaction
with the defect is less relevant.  In fact, it can be shown that the attraction
of the columnar defect must exceed a finite threshold before it can pin
a FL\cite{Bal94}. Thus the results of this section regarding the Gaussian
distribution of the FL winding angle in the presence of point impurities are
expected to hold even for the more realistic attractive columnar pins.
Perhaps not surprisingly, the main conclusion is that the pinning to point
randomness decreases entanglement.

\section{Discussion and Conclusions}
\label{concl}

Topological entanglements present strong challenges to our understanding
of the dynamics of polymers and flux lines. In this paper, we examined the windings 
of a single FL around a columnar defect. By focusing on even this simple
physical situation we were able to uncover a variety of interesting properties:
The probability distributions for the winding angles can be classified into
a number of universality classes characterized by the presence or absence
of underlying symmetries or relevant length scales.

The most ``symmetric" situation is the windings of an ideal (Brownian) walk
around a point center, described by the Cauchy distribution in Eq.~(\ref{Cauchy}).
Introduction of a finite core for the winding center (or a finite persistence length
for the walk) leads to a number of exponentially decaying distributions: If there is 
a conservation law for the walkers (reflecting boundaries or neutral defects), 
we obtain the distribution in Eq.~(\ref{refl}). Removing this conservation
(absorbing boundaries or repulsive defects) leads to the distribution in
Eq.~(\ref{abs}) whose tails decay twice as fast. Constraining the end-points
of the walker to the vicinity of the winding center leads to the related distributions
in Sec.~\ref{theorD}.

A completely new set of distributions is obtained for chiral defects, where the
walker is preferentially twisted in one direction at the winding center. These
distributions have asymmetric exponential tails, with decay constants that depend
on the degree of chirality. Strong chirality appears to lead to quite broad
distributions. A remaining challenge is to find the complete form of this 
probability distribution by solving the two dimensional diffusion equation
with moving boundary conditions.

For non-ideal walks, with a vanishing probability to return to the origin, the 
properties of the winding center are expected to be irrelevant. Both self-avoiding
walks in $d=2$ dimensions, and FLs pinned by point impurities in $d=3$, 
have wandering exponents $\nu$ larger than 1/2 and fall in this category. 
We present a scaling argument (supported by numerical data) that in this
the probability distribution has a Gaussian form in the variable $\theta/\sqrt{\ln t}$.
Not surprisingly, wandering away from the center reduces entanglement.
The characteristic width of the Gaussian form is presumably a universal 
constant that has been calculated exactly for self-avoiding walks in $d=2$.
It would be intersting to see if this constant (only estimated numerically for
the impurity pinned FLs in $d=3$) can be related to other universal properties
of the walk. Changing the correlations of impurities (and hence the exponent
$\nu$) may provide a way of exploring such dependence.

It is likely that there are other universality classes not explored in this paper.
The physical problem of FLs in superconductors provides several candidates, 
such as splayed columnar pins or a collection of pins with random strengths. 
One example, is defects with randomly changing chirality, which induce a change 
$\pm \Delta \theta$ in winding angle each time the line returns to the defect.
Such randomness is in fact irrelevant in the limit $t\to \infty$, since it does
not lead to no systematic change of the variable $x$. The induced variance
in $x$ also vanishes as
\begin{equation}
(\delta\Delta x)^2 \simeq \lim_{t\to \infty}\ln t \left(2\delta\Delta 
\theta\over \ln t \right) = 0.
\end{equation}
The underlying problem is certainly quite rich and seems to call for developing
some form of renormalization group analysis.

It is also interesting to search for circumstances in which the degree of winding
changes dramatically. To compete with the exponential tail of the winding angle
distribution, we need an energy proportional to the winding angle itself; a
non-local quantity. A possible physical realization is provided by a magnetized
directed polymer winding around a wire. A current $I$ in the wire generates
a magnetic field 
$$\vec B(\vec r) ={2 \vec I \times \vec r \over cr^2}, $$
where $c$ is the speed of light. Assuming that the polymer carries a uniform
magnetic moment density $\mu$ aligned with its backbone, then leads to an energy 
\begin{eqnarray*}
E &=& -\mu  \int_0^t d\vec r(t') \cdot\vec B(\vec r)\\
&=& -{4\pi\mu I n\over c} \equiv \epsilon n\, ,
\end{eqnarray*}
proportional to the winding number $n$. 

In the presence of an energy $\epsilon$ per winding, and assuming absorbing
boundary conditions, the partition function is given by
$$Z = \int_{-\infty}^\infty dn \exp\left(-{\epsilon n \over k_BT}\right)
{\pi \over 4 \cosh^2\left(2\pi^2n/\ln t\right)}\,,$$
where $n = \theta/2\pi$ is the winding number, $k_B$ the Boltzmann constant, 
and $T$ is the temperature. The integral ``diverges'' if  
$t \ge t^* = \exp\left(4\pi^2 k_BT/ \epsilon\right)$, a cutoff only being given by 
the maximum possible winding number, proportional to $t$. A directed path, 
although free on length scales $t<t^*$, is always coiled in the limit of $t \to \infty$. 
However, for a given length $t$, a small decrease in temperature can induce a 
sharp crossover from a free to a coiled configuration due to the exponential 
dependence of the crossover length $t^*$ on temperature. A similarly sharp 
crossover occurs for the pinning of a FL to an attractive columnar defect \cite{nel93} 
(see subsection \ref{theorD}). Both transitions are related to the number of returns 
of a random walk to the origin which scales as $\ln(t)$. 
We expect similar sharp crossovers between unentangled and braided 
configurations in other cases, such as several FLs winding around each other.

\acknowledgements
We thank A. Grosberg, Y. Kantor, P. LeDoussal, and S. Redner 
for helpful discussions.  BD is supported  by the  Deutsche 
Forschungsgemeinschaft (DFG) under Contract No.~Dr 300/1-1. MK  
acknowledges support from NSF grant number DMR-93-03667.

\end{multicols}


\begin{references}{}
\bibitem{Gen71} P. G. de Gennes, J. Chem. Phys. {\bf 55}, 572 (1971).
\bibitem{nel88} D. R. Nelson, Phys. Rev. Lett. {\bf 60}, 1973 (1988).
\bibitem{obu90} M. Rubinstein and S. P. Obukhov, Phys. Rev. Lett. 
{\bf 65}, 1279 (1990).
\bibitem{nel93} D. R. Nelson, in {\it Phase Transitions and Relaxation in 
Systems with Competing Energy Scales}, edited by T. Riste and D. C. Sherrington
 (Kluwer Academic Publishers, Dordrecht, Boston, 1993).
\bibitem{spi58} F. Spitzer, Trans. Am. Math. Soc. {\bf 87}, 187 (1958).
\bibitem{pit86} J. W. Pitman and M. Yor, Ann. Probab. {\bf 14}, 733 (1986).
\bibitem{gal87} J. F. Le Gall and M. Yor, Probab. Th. Rel. Fields {\bf 74}, 
617 (1987).
\bibitem{ito65} K. It\^o, and H. P. McKean, {\it Diffusion Processes and 
Their Sample Paths} (Springer-Verlag, Berlin, 1965).
\bibitem{edw67} S. F. Edwards, Proc. Phys. Soc. London {\bf 91}, 513 (1967).
\bibitem{wie83} F. W. Wiegel, in {\it Phase Transitions and Critical 
Phenomena}, edited by C. Domb and J. L. Lebowitz (Academic, London, 1983), 
Vol.~7.
\bibitem{dur84} R. Durett, {\it Brownian Motion and Martingales in Analysis} 
(Wadsworth, Inc., Belmont, California, 1984).
\bibitem{dup88} B. Duplantier and H. Saleur, Phys. Rev. Lett. {\bf 60}, 2343 
(1988).
\bibitem{rud88} J. Rudnick and Y. Hu, Phys. Rev. Lett. {\bf 60}, 712 (1988).
\bibitem{rud87} J. Rudnick and Y. Hu, J. Phys. A: Math. Gen. {\bf 20}, 4421 
(1987).
\bibitem{bel89} C. B\'elisle, Ann. Prob. {\bf 17}, 1377 (1989). 
\bibitem{sal94} H. Saleur, Phys. Rev. E {\bf 50}, 1123 (1994).
\bibitem{gro93} A. Grosberg and S. Nechaev, {\it Polymer Topology}, Advances
in Polymer Science {\bf 106}, 1-29 (Springer Verlag, Berlin 1993).
\bibitem{kardarrev} M. Kardar, {\it Lectures on Directed Paths in 
Random Media},  Les Houches Summer School on Fluctuating Geometries in 
Statistical Mechanics and Field Theory, August 1994 (in press, see  
cond-mat/9411022).
\bibitem{kni81} F. B. Knight, {\it Essentials of Brownian motion and 
diffusion} (American Mathematical Society, Providence, Rhode Island, 1981). 
\bibitem{HTSC} 
G. Blatter, M. V. Feigel'man, V. B. Geshkenbein, A. I. Larkin, and V.  M. Vinokur, 
Rev. Mod. Phys. {\bf 66}, 1125 (1994).
\bibitem{AmarFamily} 
J. G. Amar and F. Family, Phys. Rev. A {\bf 41}, 3399 (1990).
\bibitem{KimBrayMoore} 
J. M. Kim, A. J. Bray, and M. A. Moore, Phys. Rev. A {\bf 44}, 2345 (1991).
\bibitem{Bal94} L. Balents and M. Kardar, Phys. Rev. E {\bf 49}, 13 030 
(1994).
\end{references}
\end{document}